\documentclass[twocolumn,amsmath,aps,fleqn]{revtex4}
\usepackage{graphicx,amssymb}
\begin{document}
\newcommand{\be}{\begin{equation}}
\newcommand{\ee}{\end{equation}}
\newcommand{\bq}{\begin{eqnarray}}
\newcommand{\eq}{\end{eqnarray}}
\newcommand{\bsq}{\begin{subequations}}
\newcommand{\esq}{\end{subequations}}
\newcommand{\bc}{\begin{center}}
\newcommand{\ec}{\end{center}}
\newcommand {\R}{{\mathcal R}}
\newcommand{\al}{\alpha}
\newcommand\lsim{\mathrel{\rlap{\lower4pt\hbox{\hskip1pt$\sim$}}
    \raise1pt\hbox{$<$}}}
\newcommand\gsim{\mathrel{\rlap{\lower4pt\hbox{\hskip1pt$\sim$}}
    \raise1pt\hbox{$>$}}}

\title{Vacuum energy sequestering and cosmic dynamics}

\author{P.P. Avelino}
\email[Electronic address: ]{pedro.avelino@astro.up.pt}
\affiliation{Instituto de Astrof\'{\i}sica e Ci\^encias do Espa{\c c}o, Universidade do Porto, CAUP, Rua das Estrelas, PT4150-762 Porto, Portugal}
\affiliation{Centro de Astrof\'{\i}sica da Universidade do Porto, Rua das Estrelas, PT4150-762 Porto, Portugal}
\affiliation{Departamento de F\'{\i}sica e Astronomia, Faculdade de Ci\^encias, Universidade do Porto, Rua do Campo Alegre 687, PT4169-007 Porto, Portugal}

\date{\today}
\begin{abstract}
We explicitly compute the dynamics of closed homogeneous and isotropic universes permeated by a single perfect fluid with a constant equation of state parameter $w$ in the context of a recent reformulation of general relativity, proposed in \cite{Kaloper:2013zca}, which prevents the vacuum energy from acting as a gravitational source. This is done using an iterative algorithm, taking as an initial guess the background cosmological evolution obtained using standard general relativity in the absence of a cosmological constant. We show that, in general, the impact of the vacuum energy sequestering mechanism on the dynamics of the universe is significant, except for the $w=1/3$ case where the results are identical to those obtained in the context of general relativity with a null cosmological constant. We also show that there are well behaved models in general relativity that do not have a well behaved counterpart in the vacuum energy sequestering paradigm studied in this paper, highlighting the specific case of a quintessence scalar field with a linear potential. 
\end{abstract}
\maketitle

\section{\label{intr}Introduction}

Solving the cosmological constant problem constitutes one of the most ambitious challenges of fundamental physics \cite{Padmanabhan:2002ji}. The latest constraints \cite{Suzuki:2011hu,Anderson:2012sa,Parkinson:2012vd,Hinshaw:2012aka,Ade:2013zuv} suggest that a cosmological constant may be responsible for the observed acceleration of the universe, assuming that gravity is described by general relativity on cosmological scales. However, this interpretation of the data faces several problems: i) why is the vacuum energy density about $120$ orders of magnitude smaller than the Planck density? ii) why do we seem to live at a very special epoch where the fractional contribution of the cosmological constant to the energy density of the universe appears to be rapidly evolving from $0$ in the relatively recent past towards $1$ in the not too distant future? The answer to these questions may lie on dynamical dark energy models \cite{Zlatev:1998tr,Zlatev:1998yg,Dodelson:2001fq,Malquarti:2003hn}, finite lifetime cosmologies in which the matter and dark energy densities can be of the same order for most of the universe lifetime \cite{Caldwell:2003vq,Kallosh:2003bq,Wang:2004nm,Scherrer:2004eq,Avelino:2004vy,Barreira:2011qi}, and/or anthropic considerations \cite{Weinberg:1987dv,Garriga:1999hu,Egan:2007ht,Barreira:2011qi}.

Another related problem has to do with the fact, unlike the other fundamental interactions, general relativity is not invariant under the shifting of the Lagrangian by a constant, implying that the vacuum energy density is a source for the gravitational field in general relativity. This has been a matter of debate for many years, with some authors arguing that a satisfactory solution to the cosmological constant problem requires a modification of general relativity (see, for example, \cite{Padmanabhan:2007xy}). In \cite{Kaloper:2013zca} (see also \cite{Kaloper:2014dqa,Kaloper:2014fca}) a new mechanism was proposed which prevents the vacuum energy from acting as a gravitational source, thus providing a possible explanation for the huge discrepancy between the estimation of the vacuum energy density from quantum zero-point fluctuations and the value inferred from cosmological observations. In the context of this reformulation of general relativity the universe should be finite in space and time, with the present epoch of accelerated expansion being a transient stage before the big crunch.

In the present paper we shall investigate how the cosmological dynamics is affected by the vacuum energy sequestering mechanism. This paper is organized as follows. In Sec. \ref{sec2}, some of the key features of the theory proposed in \cite{Kaloper:2013zca} are outlined. In Sec. \ref{sec3}, we use an iterative algorithm to determine the impact of the vacuum energy sequestering mechanism on the dynamics of closed homogeneous and isotropic universes filled with a perfect fluid with a constant equation of state parameter. The results are then compared with those obtained in the context of general relativity with a null cosmological constant. In this section we also explore the implications of the vacuum energy mechanism in the context of more general models, with special emphasis to the case of a quintessence scalar field with a linear potential. We then conclude in Sec. \ref{conc}.

Throughout this paper we use units such that $8\pi G=c=1$, where $G$ is the gravitational constant and $c$ is the value of the speed of light in vacuum. We adopt the metric signature $(-,+,+,+)$.

\section{The model\label{sec2}}

Here we shall consider the action defined in \cite{Kaloper:2013zca} which yields the following equations of motion for the gravitational field
\be
G^{\mu\nu}=T^{\mu \nu} - \Lambda g^{\mu\nu}\,,
\label{einst}
\ee
where $G^{\mu\nu} \equiv R^{\mu\nu}-g^{\mu\nu}R/2$ are the components of the Einstein tensor, $g^{\mu\nu}$ are the components of the metric, $R^{\mu\nu}$  are the components of the Ricci curvature tensor, $R \equiv {R^{\mu}}_{\nu}$ is the Ricci scalar curvature, $T^{\mu \nu}$ are the components of the energy momentum tensor and $\Lambda$ is given by
\be
4 \Lambda=\langle {T^{\mu}}_{\mu} \rangle \equiv \frac{\int d^4 x {\sqrt {-g}}{T^{\mu}}_{\mu}}{\int d^4 x {\sqrt {-g}}}\,,
\label{lambda}
\ee
where $g={\rm det}(g_{\mu\nu})$ is the metric determinant. These equations of motion for the gravitational field are invariant under the transformation $T^{\mu \nu} \to T^{\mu \nu}+C g^{\mu\nu}$ where $C$ is an arbitrary real constant. Consequently, any bulk constant energy density is effectively gauged away.

\begin{figure}
\begin{center}
\includegraphics*[width=9cm]{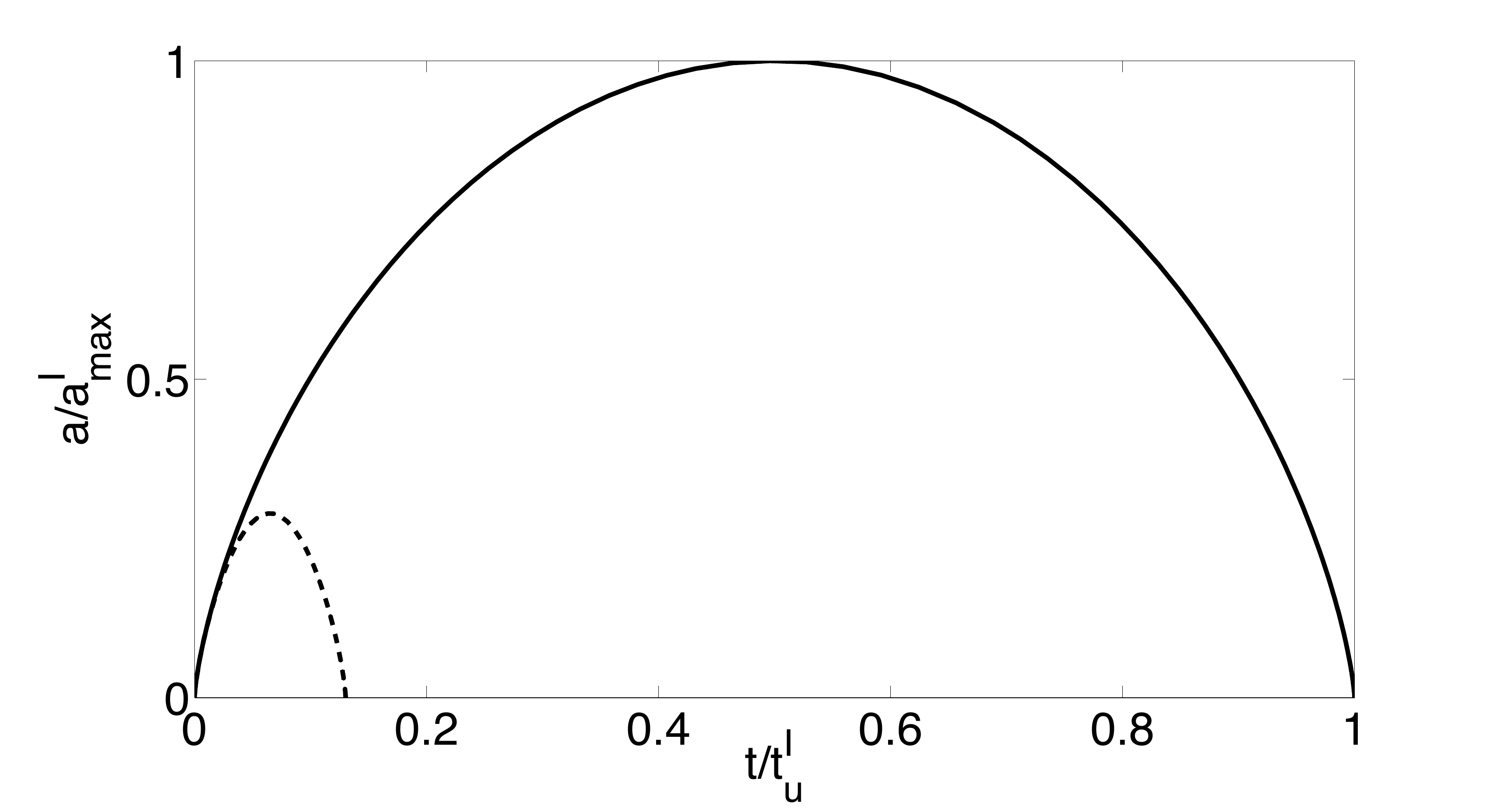}
\end{center}
\caption{\label{fig1} Evolution of the scale factor with cosmic time for a model with $w=-0.1$ considering  general relativity with a null cosmological constant (model I, solid line) and the reformulation of general relativity incorporating the vacuum energy sequestering mechanism (model II, dashed line). The maximum value of the scale factor $a_{\rm max}$ and the universe lifetime $t_{\rm u}$ in model I are both normalized to unity.}
\end{figure}

In this paper we shall consider a closed homogeneous and isotropic universe described by the Friedmann-Lemaitre-Robertson-Walker metric. The line element is given by
\be
ds^2=-dt^2+a(t)^2\left(\frac{dr^2}{1-kr^2}+r^2\left(d\theta^2+\sin^2 \theta d\phi^2\right)\right)\,,
\label{metric} 
\ee
where $t$ is the physical time, $(r,\theta,\phi)$ are comoving spherical coordinates and $k>0$ is the constant curvature of the 3-dimensional space. In a homogeneous and isotropic spacetime the energy-momentum tensor of the background source must have a perfect fluid form
\be
T^{\mu \nu}=(\rho+p)u^\mu u^\nu+pg^{\mu\nu} \,,
\ee
where $\rho$ is the energy density, $p$ is the pressure and $u^\mu$ are the components of the four-velocity of the fluid ($u^0=-1$ and $u^i=0$) in comoving coordinates. The trace of the energy-momentum tensor is
\be
{T^{\mu}}_{\mu}=-\rho+3p=\rho(3w-1)\,,
\ee
where $w=p/\rho$ is the equation of state parameter. 

\begin{figure}
\begin{center}
\includegraphics*[width=8.5cm]{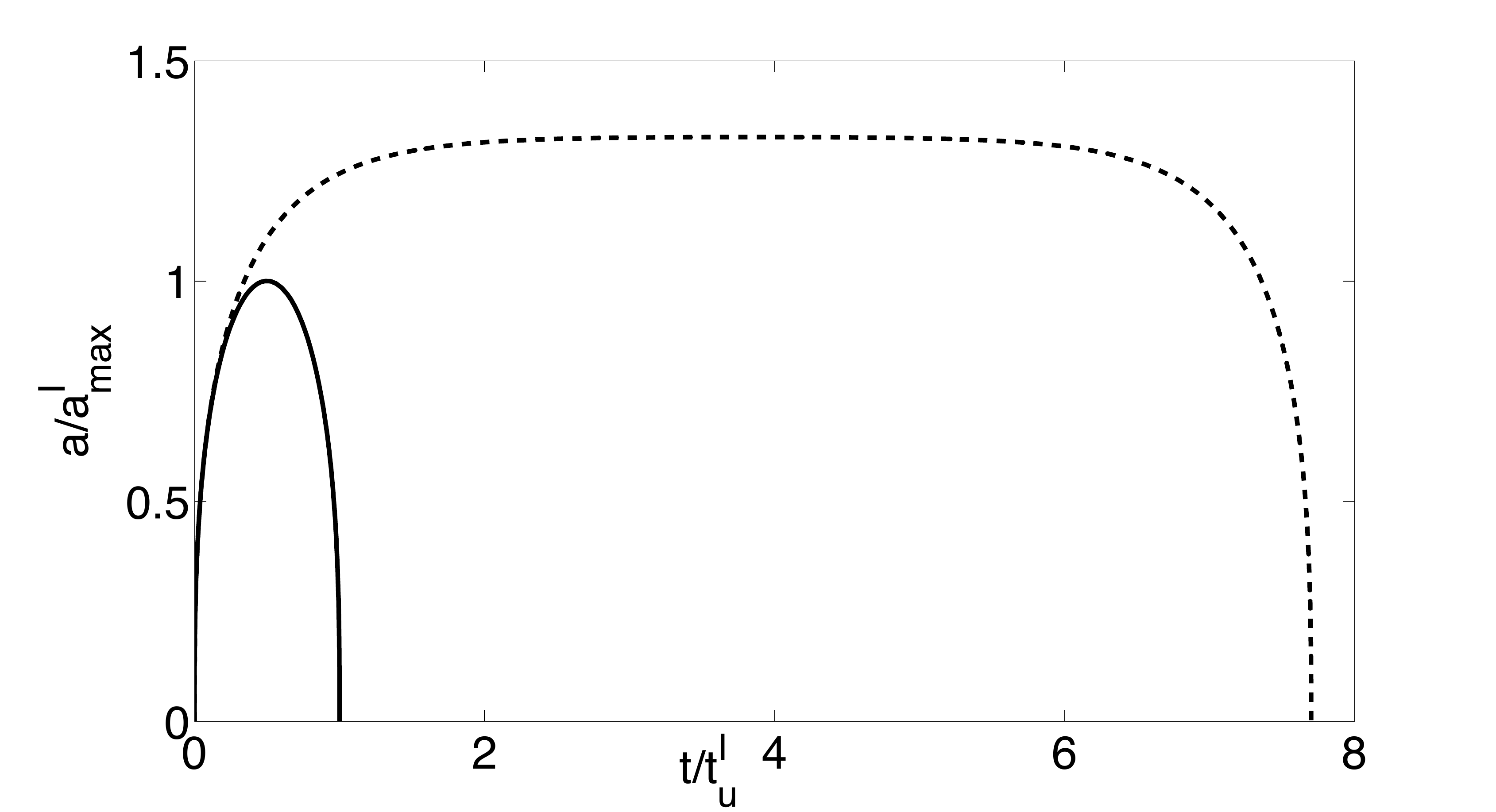}
\end{center}
\caption{\label{fig2} The same as in Fig. \ref{fig1} but with $w=0.9$}
\end{figure}

The dynamics of the universe can be obtained from Eqs. (\ref{einst}) and (\ref{lambda}), considering the metric given by  Eq. (\ref{metric}). The equations of motion are given by
\bq
H^2&=&\frac{\rho}{3} +\frac{\Lambda}{3}- \frac{k}{a^2}\,, \label{hubble} \\
\frac{\ddot a}{a}&=&-\frac{\rho(1+3w)}{6}+\frac{\Lambda}{3}\,, \label{acc}
\eq
with
\be
{T^{\mu}}_{\mu}=\rho(3w-1)=4\Lambda - 6\left(H^2+\frac{\ddot a}{a}+\frac{k}{a^2}\right)\,.
\label{lambda2}
\ee
Eqs. (\ref{lambda}) and (\ref{lambda2}) then imply the following restriction on the overall dynamics of the universe
\be
\left\langle H^2+\frac{\ddot a}{a}+\frac{k}{a^2}\right\rangle = 0\,.
\label{average}
\ee
This restriction is a direct consequence of the vacuum energy sequestering mechanism and it is not present in general relativity.

\section{Cosmic dynamics\label{sec3}}

\subsection{Constant $w$ models}

Here we use an iterative algorithm to determine the cosmological evolution of universes permeated by a single perfect fluid with a constant equation of state parameter $w$ (in which case $\rho \propto a^{-3(w+1)}$). Starting from the $w=0.3$ model the algorithm calculates the value of $\Lambda$ for each $w$ (considering fixed positive or negative steps of $\Delta w$). For each $w$ the evolution of the universe is computed and the value of $\Lambda$ is estimated iteratively using 
\be
\Lambda= \frac14 \langle {T^{\mu}}_{\mu} \rangle = \frac{3w-1}{4} \frac{\int dt a^3 \rho}{\int dt a^3}\,,
\label{lambda1}
\ee
Note that for $w=1/3$ the universe dynamics is not affected by the vacuum energy sequestering mechanism  ($\Lambda=0$ in that case). However, this is no longer the case for $w \neq 1/3$ ($\Lambda<0$ for $w < 1/3$ and $\Lambda>0$ for $w > 1/3$). In the iterative process $\Lambda=0$ is taken as an initial guess for $0.3-3\Delta w <w<0.3+3\Delta w$. For $w<0.3-3\Delta w$ and $w>0.3+3\Delta w$ the initial condition for $\ln |\Lambda|$ is obtained from the previous two $w$ steps using a linear fit. For large values of $w$ (larger than 0.7), the iterative procedure does not always converge and a few additional constraints are necessary in order to ensure convergence. The numerical results presented in this section have considered $w_{\rm min}=-0.32$, $w_{max}=0.90$ and  $\Delta w=0.01$. 

Fig. \ref{fig1} shows the evolution of the scale factor with cosmic time for a model with $w=-0.1$ considering: i) general relativity with a null cosmological constant (model I, solid line) ii) the reformulation of general relativity incorporating the vacuum energy sequestering mechanism (model II, dashed line). The maximum value of the scale factor $a_{\rm max}$ and of the universe lifetime $t_{\rm u}$ in model I are both normalized to unity. Fig. \ref{fig1} illustrates the large impact that this vacuum energy sequestering mechanism has on the the dynamics of the universe for values of $w$ not too far from $-1/3$. Close to this limit, the first and the last term on the r.h.s. of Eq.  (\ref{hubble}) have a similar evolution with the scale factor, and, consequently, in model I the stage with $H \sim 0$ corresponds to a very large variation of the scale factor. This implies that the introduction of the negative $\Lambda$ term (in model II) has a dramatic effect on the dynamics of the universe for values of $w$ larger but close to $-1/3$, leading to much smaller universe lifetimes $t_{\rm u}$ and maximum values of the scale factor $a_{\rm max}$ compared to model I, as illustrated in Fig. \ref{fig1}.

\begin{figure}
\begin{center}
\includegraphics*[width=9cm]{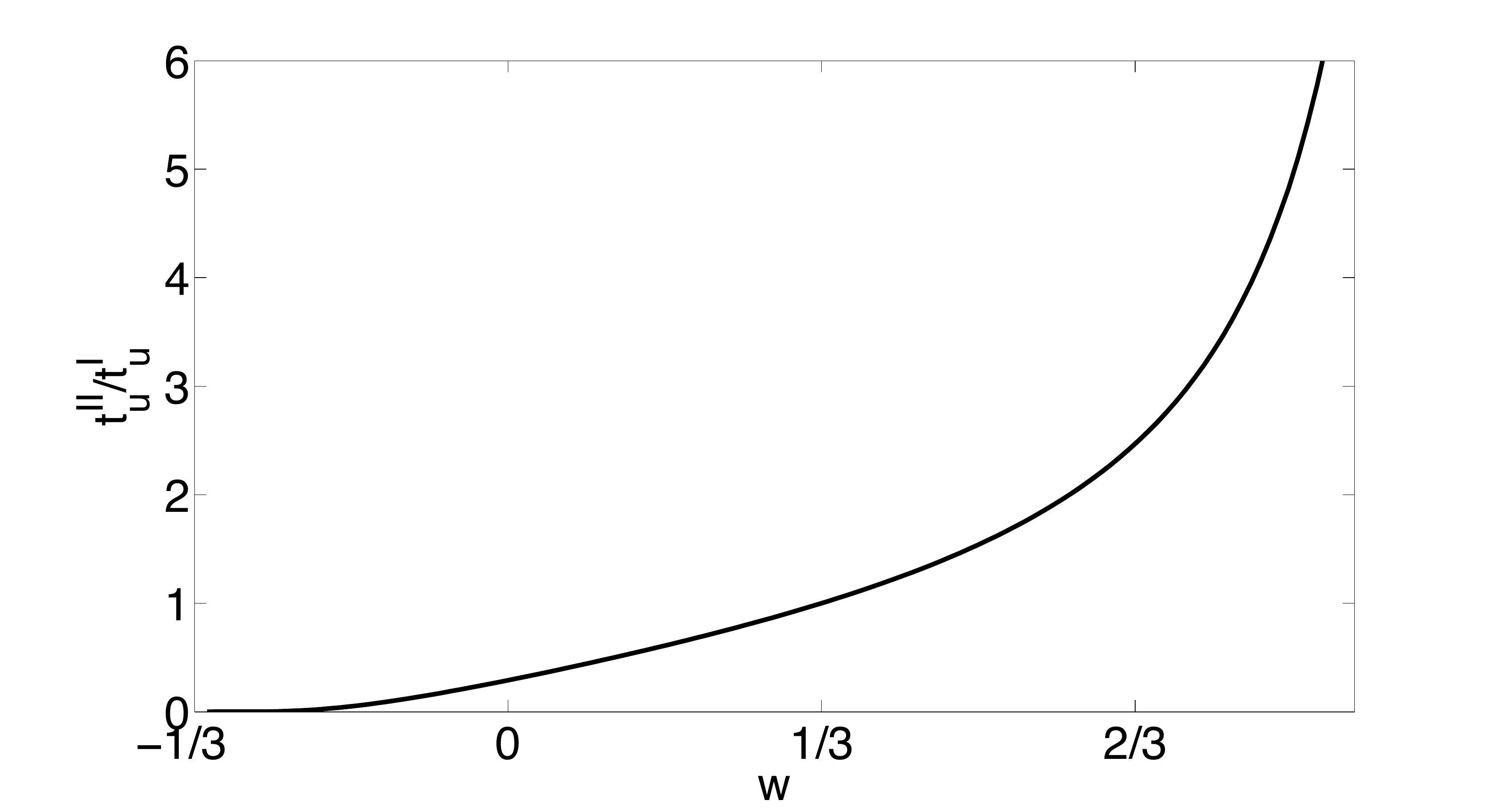}
\end{center}
\caption{\label{fig3} Ratio between the universe lifetimes in models II and I as a function of $w$.}
\end{figure}

Fig. \ref{fig2} is analogous to Fig. \ref{fig1} but now considering $w=0.9$.  Fig. \ref{fig2} illustrates the very large impact that the vacuum energy sequestering mechanism has on the the dynamics of the universe for $w \to 1^{-}$. Although $\int dt a^3 \rho$ diverges near $a=0$ for $w=1$ (note that $\rho \propto a^{-6}$ and $a \propto t^{1/3}$ for $w=1$ near the big bang, an epoch in which curvature of the universe has no impact on its dynamics), there is a well behaved model II solution in the  $w \to 1^{-}$ limit. Close to that limit, the universe may spend an arbitrary large amount of time in a quasi static state with $H \sim 0$ and ${\ddot a}/a \sim 0$, which weighs down the contribution of the phase close to $a=0$ in the calculation of $\Lambda$. Fig. \ref{fig2} illustrates this, showing that the universe lifetime is much larger for model II than for model I.

\begin{figure}
\begin{center}
\includegraphics*[width=9cm]{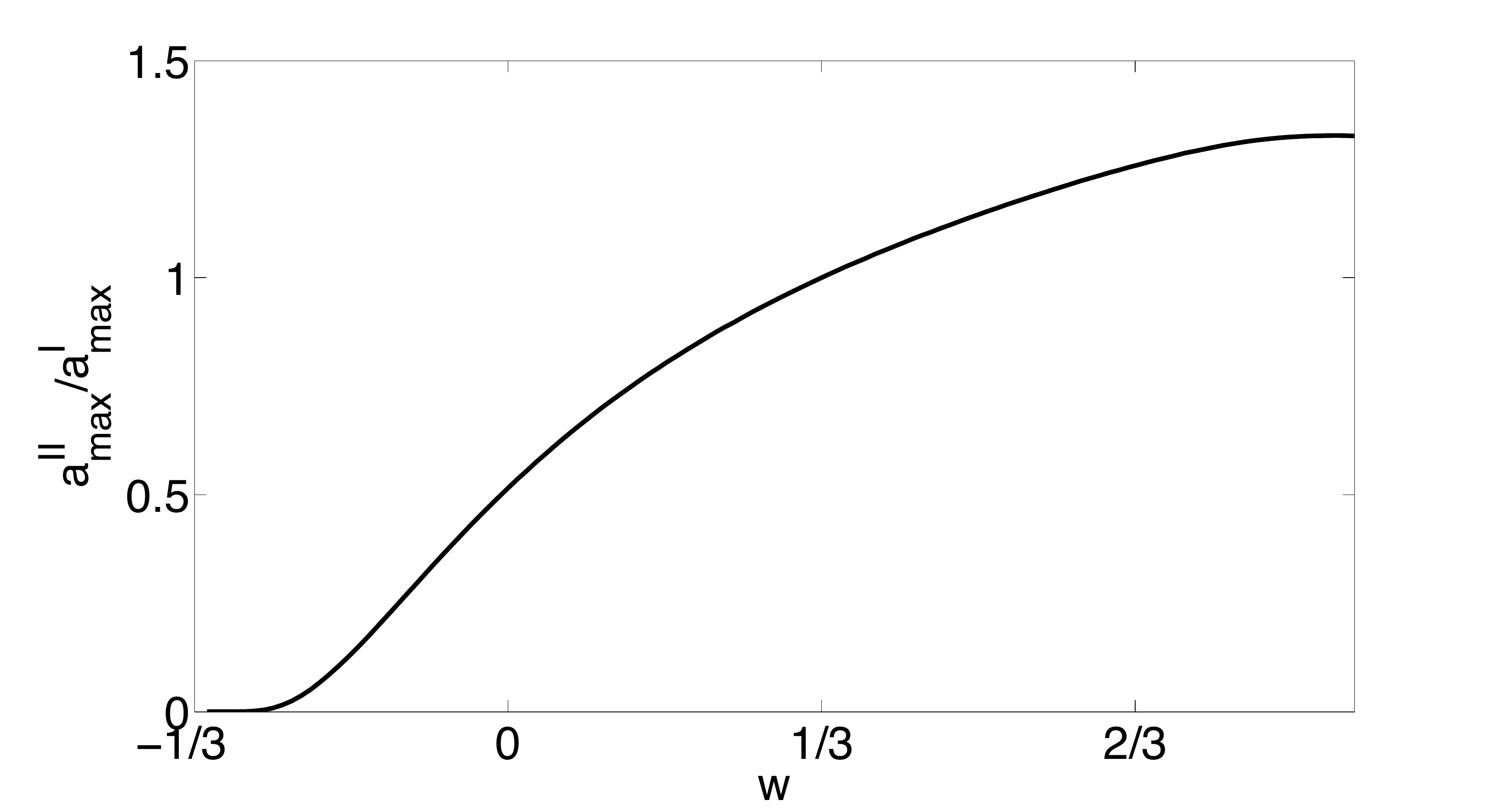}
\end{center}
\caption{\label{fig4} Ratio between the maximum values of the scale factor in models II and I as a function of $w$.}
\end{figure}

Fig. \ref{fig3} shows the ratio between the universe lifetimes in models II and I ($t_{\rm u}^{\rm II}/t_{\rm u}^{\rm I}$) as a function of $w$. As previously discussed, this ratio tends to zero in the $w \to -1/3^{+}$ limit and to $\infty$ in the $w \to 1^{-}$ limit. Except for the $w=1/3$ case, the reformulation of general relativity proposed in \cite{Kaloper:2013zca} to prevent the vacuum energy from sourcing the gravitational field has a significant impact on the dynamics of the universe. This can also be seen in Fig. \ref{fig4} which shows the ratio between the maximum values of the scale factor in models II and I ($a_{\rm max}^{\rm II}/a_{\rm max}^{\rm I}$) as a function of $w$. As expected, this ratio becomes very small in the $w \to -1/3^{+}$ limit and tends to a constant in the $w \to 1^{-}$ limit.

\begin{figure}
\begin{center}
\includegraphics*[width=9cm]{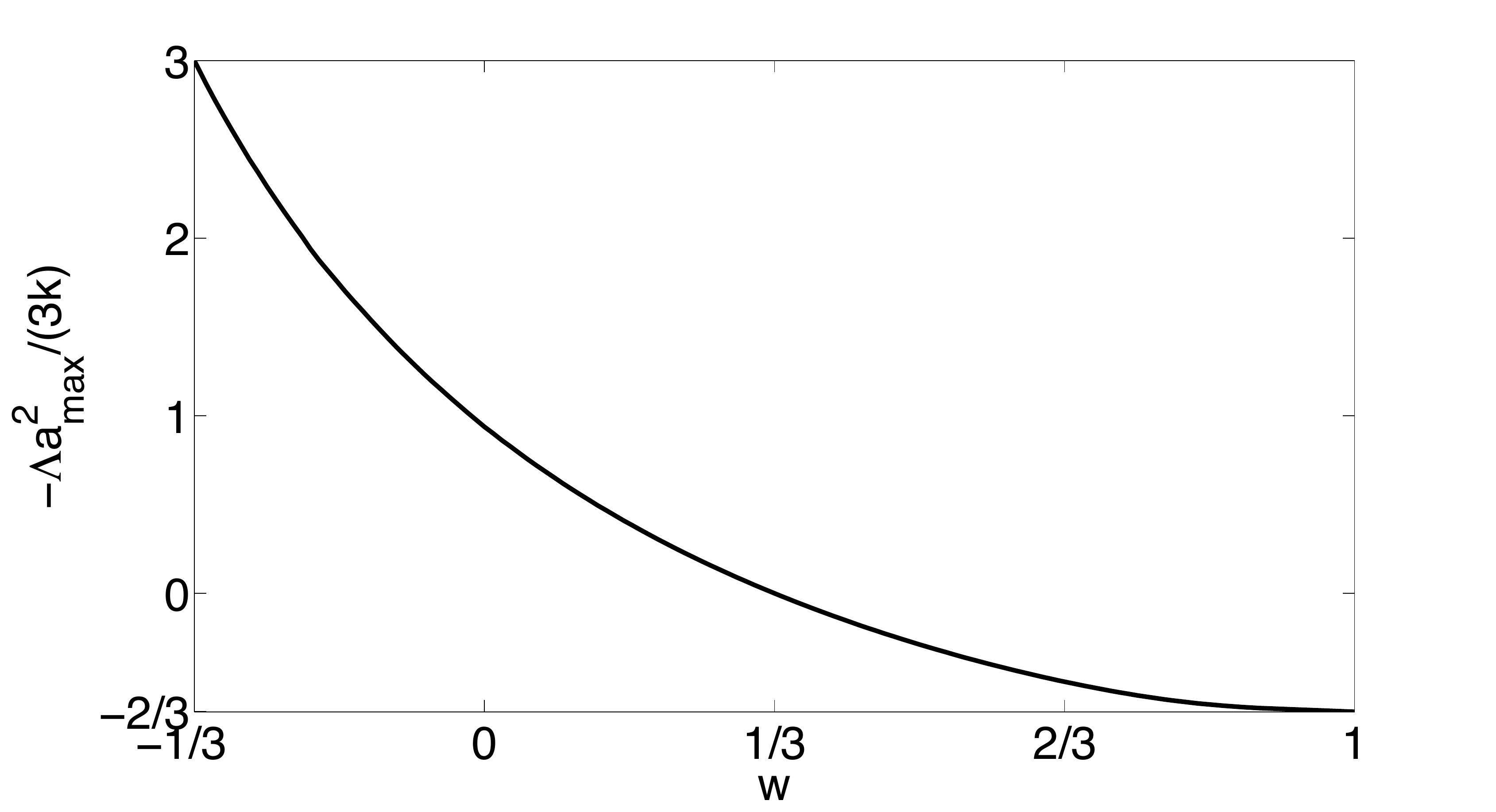}
\end{center}
\caption{\label{fig5} Ratio between $\Lambda/3$ and $-k/a_{\rm max}^2$ as a function of $w$ (model II).}
\end{figure}

\begin{figure}
\begin{center}
\includegraphics*[width=9cm]{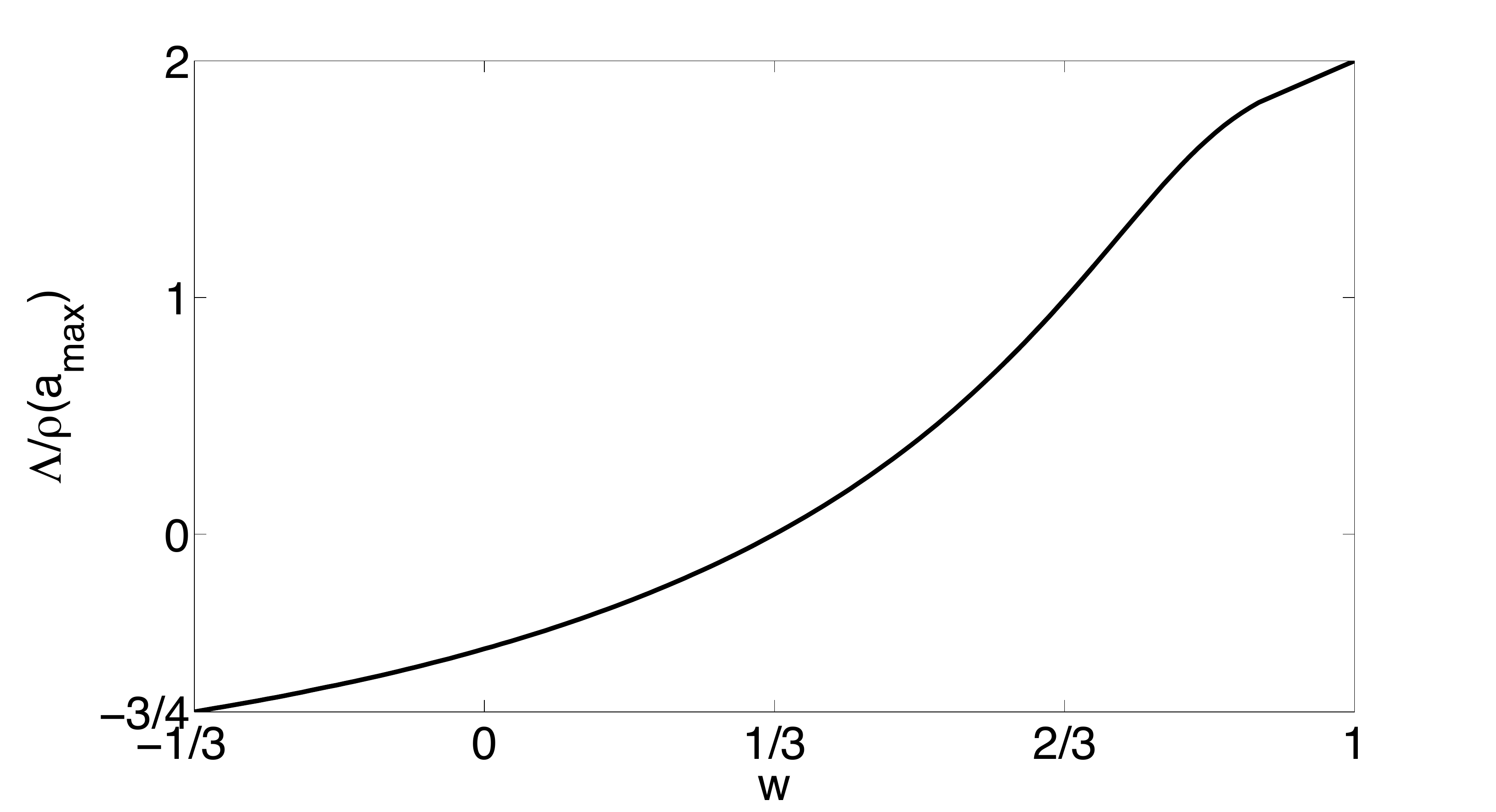}
\end{center}
\caption{\label{fig6} Ratio between $\Lambda$ and the minimum density (obtained for $a=a_{\rm max}$) as a function of $w$ (model II).}
\end{figure}

Fig. \ref{fig5} shows the ratio between the last two terms on the r.h.s. Eq. (\ref{hubble}) for $a=a_{\rm max}$ ($-\Lambda a^2_{\rm max}/(3k)$) in model II as a function of $w$ while the ratio between $\Lambda$ and the minimum density $\rho(a_{\rm max})$ for the same model is plotted in Fig. \ref{fig6} also as a function of $w$. We conclude from the plots that $|\Lambda|$ is never larger than a few times the minimum density for this family of models. We shall see in the next section that this is not always the case in a more general framework.

The ratios shown in Figs. \ref{fig5} and \ref{fig6} can be found analytically in the $w \to 1^{-}$ limit by taking into account that, in this limit,
\bq
H^2&=&\frac{\rho}{3} + \frac{\Lambda}{3}-\frac{k}{a^2}=0\,,\\  \label{eq2max}
\frac{\ddot a}{a}&=&-\frac{2}{3}\rho+\frac{\Lambda}{3}=0\,, \label{eq2max}
\eq
for $a=a_{\rm max}$. Eq.  (\ref{eq2max}) implies that $\rho(a_{\rm max}) = \Lambda/2$. Substituting in Eq.  (\ref{eq2max}) one obtains $-\Lambda a^2_{\rm max}/(3k)=-2/3$.

The value of $-\Lambda a^2_{\rm max}/(3k)$ may also be calculated analytically in the $w \to -1/3^{+}$ limit. In this limit $\rho \propto a^{-2}$ and $\Lambda<0$. Consequently, Eq. (\ref{hubble}) may be written as 
\be
H^2=\frac{\Lambda}{3}\left(1-\frac{a_{\rm max}^2}{a^2}\right)\,,
\label{weqm13}
\ee
where we have taken into account that $H=0$ for $a=a_{\rm max}$ thus implying that
\be
\frac{\rho(a_{\rm max})}{3} - \frac{k}{a_{\rm max}^2}=-\frac{\Lambda}{3}\,.
\ee
The solution to Eq. (\ref{weqm13}) is given by
\be
a=a_{\rm max} \sin\left({\sqrt{-\Lambda/3}t}\right)\,,
\ee
and the average value of $\rho$ may be calculated as 
\be
\langle \rho \rangle =a_{\rm max}^2  \rho(a_{\rm max})\frac{\int dt a}{\int dt a^3}=\frac{3}{2}\rho(a_{\rm max})\,,
\ee
taking into account that $\rho=\rho(a_{\rm max})(a_{\rm max}/a)^2$ in the $w \to -1/3^{+}$ limit.
Using Eqs. (\ref{hubble}), (\ref{acc}) and (\ref{average}) one finds that $\langle \rho \rangle = -2 \Lambda$, in the $w \to -1/3^{+}$ limit, implying that $\rho(a_{\rm max}) = -4 \Lambda/3$. We may now conclude that, in this limit, 
\be
0=H^2=\frac{\rho}{3} +\frac{\Lambda}{3}- \frac{k}{a^2}=-\frac{\Lambda}{9}- \frac{k}{a^2}\,,
\ee
for $a=a_{\rm max}$, which implies that $-\Lambda a^2_{\rm max}/(3k)=3$.

In Figs. (\ref{fig5}) and (\ref{fig6}) these analytical constraints have been used to extend the results to the $w \to -1/3^+$ and $w \to 1^-$ limits.

\subsection{Scalar field with a linear potential}

Here we shall consider a homogeneous and isotropic universe filled with matter and a standard quintessence scalar field $\phi$. In the context of general relativity and in the absence of a cosmological constant this model is fully described by the equations
\bq
H^2&=& \frac13 \left(\rho_m +{\dot \phi}^2/2-V(\phi)\right)- \frac{k}{a^2}\,, \label{hubble1}\\
{\ddot \phi}&+&3H{\dot \phi}=-\frac{dV}{d \phi}\,, \label{sfield}
\eq
where $\rho_m \propto a^{-3}$ is the matter density. Here we shall assume that the scalar field potential $V(\phi)$ is a linear function of $\phi$, namely 
\begin{equation} 
V(\phi) = V_0 + \frac{dV}{d\phi} \left(\phi-\phi_0\right)\, , 
\label{eq6}
\end{equation} 
where $|{dV}/{d\phi}|$ is a constant and the subscript `$0$' means that the variables are to be evaluated at the present time $t_0$ (see \cite{Avelino:2004vy} for more details on this model in the context of general relativity, including a possible solution to the coincidence problem).

In the dark energy dominated era the dark energy scalar field $\phi$ is constrained to be in a slow-roll regime with
\begin{equation} 
w_\phi=\frac{{\dot \phi}^2/2-V(\phi)}{{\dot \phi}^2/2+V(\phi)} \sim -1\,,
\label{eq4}
\end{equation}
and
\be
3H{\dot \phi}\sim-\frac{dV}{d \phi}\,.
\ee
In this regime the evolution of $\phi$ is very slow and the main contribution to the energy density of the universe comes from $V(\phi)$. In this phase the kinetic energy of the scalar field increases very slowly due to the corresponding very slow decrease of $V$. However, no matter how small the value of $|\frac{dV}{d \phi}| > 0$ is, at some point the slow-roll regime ends, $V(\phi)$ turns negative and the universe collapses with the energy density becoming dominated by the kinetic energy density of the scalar field $\phi$ until the big crunch.

The conclusion that the universe eventually collapses and that the late time evolution of the universe is dominated by the kinetic energy of the scalar field $\phi$ remains valid even if one allows for an additional finite $\Lambda$ contribution associated to the vacuum energy sequestering mechanism (this has also been shown in \cite{Kaloper:2014fca}). Also, near the big crunch the universe is nearly flat and the curvature may also be neglected. 

Eq. (\ref{sfield}) implies that
\be
H^2 \phi'' + a \left(\frac{\ddot a}{a}+2H^2\right)\phi'=-\frac{dV}{d \phi}\,,
\ee
where a prime denotes a derivative with respect to $\ln a$. Neglecting the curvature and $\Lambda$ terms in Eq. (\ref{hubble}), setting $w=1$ in Eq. (\ref{acc}) and taking into account that $\rho \propto a^{-6}$ for $w=1$ one obtains that 
\be
a^{-6} \phi'' = - C_1 \frac{dV}{d \phi}\,, \label{sfieldweq1}
\ee
where $C_1$ is a positive constant. The solution to Eq. (\ref{sfieldweq1}) is then given by
\be
\phi=- \frac{C_1}{30} \frac{dV}{d \phi}a^6 + C_2+C_3 \ln a\,.
 \label{solsfieldweq1}
\ee
Hence, the values of $\phi$ and $V(\phi)$ display only a relatively slow change with $a$ as the universe approaches the big crunch. On the other hand, the kinetic energy of the scalar field is increasing proportionally to $a^{-6}$, driving the value of $w$ closer and closer to unit. This results in an infinite value for $\Lambda$ if the vacuum energy mechanism is applied, thus invalidating this quintessence model as a viable cosmological scenario (in \cite{Kaloper:2014fca} the authors did not consider the possibility of a divergent $\Lambda$ which resulted in a different conclusion). Note that this only occurs for models where $w$ tends to unity sufficiently fast at the big crunch (or at big bang) so as to make the integral in Eq. (\ref{lambda}) diverge, which does not happen in general. 

\section{\label{conc}Conclusions \label{conc}}

In this paper we have quantified the impact of a recently proposed vacuum energy sequestering mechanism on the background evolution of the universe, using both analytical and numerical analysis. We confirmed that, in general, this mechanism significantly modifies the dynamics of the universe with respect to the cosmological dynamics obtained in the context of general relativity with a null cosmological constant. We have shown that in some cases, in particular for values of $w$ close to $-1/3$ and $1$, the dynamical changes can be dramatic. We have also shown that there are well behaved quintessence models in the context of general relativity which do not have a well behaved counterpart in the vacuum energy sequestering paradigm studied in this paper. We have highlighted the particular case of a quintessence scalar field with a linear potential, which has been suggested as a possible solution to the coincidence problem in the context of general relativity. 

\begin{acknowledgments}

This work was supported by Funda{\c c}\~ao para a Ci\^encia e a Tecnologia (FCT) through the Investigador FCT contract of reference IF/00863/2012 and POPH/FSE (EC) by FEDER funding through the program "Programa Operacional de Factores de Competitividade - COMPETE. P.P.A. is also partially supported by grant PTDC/FIS/111725/2009 (FCT).

\end{acknowledgments}


\bibliography{Lambda}

\end{document}